\documentstyle[preprint,aps,tighten]{revtex}
\newcommand{\ben}{\begin{displaymath}}
\newcommand{\een}{\end{displaymath}}
\newcommand{\be}{\begin{equation}}
\newcommand{\ee}{\end{equation}}
\newcommand{\bea}{\begin{eqnarray}}
\newcommand{\eea}{\end{eqnarray}}
\newcommand{\eqn}[1]{\label{#1}}
\newcommand{\eq}[1]{Eq.~(\ref{#1})}
\newcommand{\eqs}[1]{Eqs.\ (\ref{#1})}

\newcommand{\bP}{\bar{P}}
\newcommand{\tPhi}{\tilde{\Phi}}
\newcommand{\bpsi}{\bar{\psi}}
\newcommand{\bPsi}{\bar{\Psi}}
\newcommand{\bPhi}{\bar{\Phi}}

\newcommand{\bfP}{{\bf P}}

\newcommand{\tK}{\tilde{K}}
\newcommand{\tG}{\tilde{G}}

\begin{document}
\draft
\title{Perturbation theory for bound states and resonances where potentials and propagators have arbitrary energy dependence}
\author{A. N. Kvinikhidze$^{1,2,}$\footnote{On leave from Mathematical Institute
of Georgian Academy of Sciences, Tbilisi, Georgia.} and B. Blankleider$^{2}$}
\address{$^{1}$Department of Physics and Astronomy, University of Manchester,
Manchester M13 9PL, United Kingdom}
\address{$^{2}$Department of Physics, The Flinders University of South
Australia, Bedford Park, SA 5042, Australia}
\date{\today}
\maketitle
\begin{abstract}

Standard derivations of ``time-independent perturbation theory'' of quantum
mechanics cannot be applied to the general case where potentials are energy
dependent or where the inverse free Green function is a non-linear function of
energy. Such derivations cannot be used, for example, in the context of
relativistic quantum field theory. Here we solve this problem by providing a
new, general formulation of perturbation theory for calculating the changes in
the energy spectrum and wave function of bound states and resonances induced by perturbations to the Hamiltonian. Although our derivation is valid for
energy-dependent potentials and is not restricted to inverse free Green
functions that are linear in the energy, the expressions obtained for the energy
and wave function corrections are compact, practical, and maximally similar to
the ones of quantum mechanics. For the case of relativistic quantum field
theory, our approach  provides a direct covariant way of obtaining corrections to  bound and resonance state masses, as well as to
wave functions that are not in the centre of mass frame.

\end{abstract}



\section{Introduction}

There is a growing interest in calculations, within a covariant quantum field
theory framework, of changes in the properties of bound states and resonances
induced by small perturbations in the interaction Hamiltonian.  The
four-dimensional Bethe-Salpeter equation and its various three-dimensional
reductions (so-called quasi-potential equations) are the most popular tools in
this respect.  A current example is the Nambu Jona-Lasinio (NJL) model where the
nucleon is described in terms of three relativistic quarks interacting via
contact potentials, and where meson exchange provides an important perturbative
correction \cite{Ishii}. Another example is provided by relativistic
calculations of hadronic atoms where the strong interaction perturbs the Coulomb
bound state \cite{Ivanov,Sazdjian}, and yet another by various other corrections
to relativistic calculations of electromagnetic bound states \cite{Faustov}.

The perturbation problem involved in such covariant calculations can be
formulated as follows. Denoting the total four-momentum of the system by $P$,
one would like to determine the bound state solution of the equation
\be
\left[ G_0^{-1}(P)-K_0(P)-K_1(P)\right]\Psi=0      \eqn{BS_Psi}
\ee
where $K_1(P)$ is a perturbation to the unperturbed kernel $K_0(P)$, and where
it is assumed that the unperturbed Green function $G_u(P)$, defined as the
solution to the equation
\be
G_u(P) = G_0(P) + G_0(P)K_0(P)G_u(P), \eqn{G_u}
\ee
is known completely.\footnote{For simplicity of presentation we generally do not show spin or relative momentum variables; similarly, identical particle factors and
all sums and integrals over intermediate state variables are suppressed.} Thus we seek the mass $M$ and wave function $\Psi$ such
that \eq{BS_Psi} with $P^2=M^2$ is satisfied.  A consequence of the complete
knowledge of $G_u(P)$ is that the mass spectrum $M_u^n$ ($n=1,2,3,\ldots$)
and corresponding wave functions $\Phi_n$ of the unperturbed
equation
\be
\left[ G_0^{-1}(P)-K_0(P)\right]\Phi_n=0 \eqn{BS_Phi}
\ee
where $P^2=(M_u^n)^2$, are known.
\smallskip

%It is important to note that in quantum field theory the reverse is
%not true - the knowledge of all the $M_u^n$ and wave functions $\Phi_n$ does
%not determine $G_u(P)$ completely. For this reason we stress that it is
%$G_u(P)$ which is assumed to be known here.

The task of solving \eq{BS_Psi} by expressing the mass $M$ and wave function
$\Psi$ as a perturbation series with respect to $K_1$ is a problem whose
solution is well-known in the corresponding context of non-relativistic quantum
mechanics (given by so-called time-independent perturbation
theory). Unfortunately the (textbook) derivation used to obtain the quantum
mechanical result is restricted to the case where the inverse free Green
function $G_0^{-1}(P)$ is linearly dependent on energy $P_0$ and where the
unperturbed kernel $K_0$ is an energy-independent Hermitian operator. Although
these restrictions lead to the closure and orthonormality conditions
\be
\bPhi_n\Phi_m = \delta_{nm},\hspace{2cm} \sum_n \Phi_n \bPhi_n = 1, \eqn{Phi_qm}
\ee
which are crucial for the derivation of time-independent perturbation theory,
they are not valid in the Bethe-Salpeter case.  Indeed none of these
restrictions are required in the context of a covariant field theoretic
approach. In this paper we therefore present a new solution to the
perturbation problem which is valid for any form of $G_0^{-1}(P)$ and $K_0(P)$;
in particular, our solution is valid for the case of covariant field theoretic
approaches where $G_0^{-1}(P)$ depends nonlinearly on $P_0$ and where $K_0(P)$ can
be energy ($P_0$) dependent.
Our solution, given in \eq{Mdiff} and \eq{Psi} for the nondegenerate case, and
in \eq{Mdiff_deg} and \eq{Psi_deg} for the degenerate case, expresses the mass
$M$ of the bound state or resonance and the corresponding wave function $\Psi$
in terms of compact expressions that take into account the perturbation term
$K_1$ to any order. At the same time, our formulation allows us to write the
perturbation series for both $M$ and $\Psi$, up to any order, in a
straightforward way which is maximally close to the analogous quantum mechanical
formulation. A further important aspect of our approach is that it is manifestly
covariant. This feature enables the direct use of the perturbation series for
$\Psi$ also in cases where the bound state or resonance is not at rest. In this way the more involved
approach of Lorentz boosting wave functions calculated perturbatively  in the rest frame,
can be avoided.
As such, our approach to the perturbation problem where no
restriction is put on the energy dependence of kernels and inverse free Green
functions, may
provide some important advantages over previous formulations
\cite{Lepage,Bodwin,Ivanov}.

\section{Perturbation theory}
\subsection{Basic equations}

In this paper we use the framework of relativistic quantum field theory to
illustrate our approach to perturbation theory. Although this is done partly for
presentational purposes -- it is a particular case where the kernel is energy
dependent and where the inverse Green function is non-linearly dependent on
energy, it is also a particularly topical case, as discussed in the
Introduction. On the other hand, we emphasize that our approach to perturbation
theory does not depend on the particular theoretical framework in which the
bound state problem is set -- it can be that of non-relativistic quantum
mechanics, relativistic quantum field theory, three-dimensional relativistic
quasi-potential equations, etc. Similarly, our approach does not depend on the
functional form taken by the energy dependence of either the kernel or the
inverse free Green function. All we need to assume is the usual overall
structure of the dynamical equations involved, as exemplified by \eq{BS_Psi} and
\eq{G_u}. 

We thus consider the Green function
\be
G(P)=G_0(P)+G_0(P)K(P)G(P),         \eqn{G}
\ee
where $P$ is the total four-momentum, $G_0$ is the fully disconnected part of
$G$, and where the kernel $K$ consists of a part $K_0$ for which the
corresponding Green function is known, and a small part $K_1$ which can be
treated as a perturbation. Thus
\be
K(P)=K_0(P)+K_1(P),  \eqn{K}
\ee
and it is assumed that the unperturbed Green function $G_u(P)$ has been
previously determined by solving \eq{G_u}. We are interested in the case where 
$G_u(P)$ has a pole corresponding to a bound or resonance state.
Thus we can write
\be
G_u(P)=\frac{i\Phi(P)\bPhi(P)}{P^2-M_u^2} + G_u^b(P) \eqn{G_u_pole}
\ee
where the wave functions $\Phi(P)$ and $\bPhi(P)$, the unperturbed bound state
mass $M_u$, and the background term $G_u^b(P)$ are all assumed to be
known.\footnote{Here, for simplicity, we assume that the bound state is
nondegenerate - the degenerate case will be considered in detail in the next
subsection. Also, here and elsewhere, all references to a ``bound state'' should
be understood to include the case of a ``resonance state''.  }  In this respect it is worth noting that the pole term of \eq{G_u_pole} is separable with respect to initial and final state variables, thus for a two-body system $\bPhi(P)\equiv\bPhi(P,p)$ is a function of the
initial relative momentum $p$ while $\Phi(P)\equiv\Phi(P,p')$ is a function of the final relative momentum $p'$.
Note also, that as
$P\rightarrow \bP_u$, where $\bP_u$ is any four-vector such that $\bP_u^2=M_u^2$,
the wave functions $\Phi(P)$ and $\bPhi(P)$ must reduce to the respective 
solutions of the bound state equations
\be
\Phi(\bP_u)= G_0(\bP_u)K_0(\bP_u)\Phi(\bP_u)\hspace{1cm};\hspace{1cm}
\bPhi(\bP_u)= \bPhi(\bP_u) K_0(\bP_u)G_0(\bP_u) .  \eqn{Phi}
\ee
Although $\Phi(\bP_u)$ and $\bPhi(\bP_u)$ are therefore specified as the
solutions of the above bound state equations, for momenta $P$ not on the mass shell, $P^2\ne M_u^2$, there is no unique way to define $\Phi(P)$ [and therefore $\bPhi(P)$] since any definition can be adopted in \eq{G_u_pole} with an appropriate redefinition of the background term $G_u^b(P)$. Here we shall choose $\Phi(P)$ to be a Lorentz covariant function of the total momentum $P$, the relative momenta, and the spinor indices of the constituents (i.e.\ $\Phi(P)$ is covariant under the simultaneous transformation of all these variables). The way to construct such a $\Phi(P)$ will be discussed below. Since the full unperturbed Green function $G_u(P)$ is a Lorentz covariant function of its variables from the outset, the Lorentz covariance of the background term $G_u^b(P)$ is therefore assured.

Once the perturbation $K_1$ is included, the mass $M_u$ will shift to the
physical value $M$ and $\Phi(P)$ will modify to the wave function
$\Psi(P)$ where
\be
G(P)=\frac{i \Psi(P) \bPsi(P)}{P^2-M^2} + \hat{G}^b(P).     \eqn{G_pole}
\ee
The wave functions $\Psi(P)$ and $\bPsi(P)$ are likewise assumed to be covariant functions which reduce in the limit $P\rightarrow \bP$, where
$\bP^2=M^2$, to the respective solutions of the bound state
equations
\be
\Psi(\bP)= G_0(\bP)K(\bP)\Psi(\bP),\hspace{1cm}\mbox{and}
\hspace{1cm}
\bPsi(\bP)= \bPsi(\bP) K(\bP)G_0(\bP).
\ee

To write a perturbation series for $G$, we express $G$ in terms of the known
unperturbed Green function $G_u$ through the equation
\be
G(P)=G_u(P)+G_u(P)K_1(P)G(P),           \eqn{GG_u}
\ee
which follows from the fact that $G^{-1}= G_0^{-1}-K$ and $G_u^{-1}=
G_0^{-1}-K_0$.  By iterating \eq{GG_u} we obtain a perturbation series for
$G(P)$ with respect to the perturbation $K_1(P)$. What appears more difficult is
to find a corresponding perturbation series for the mass $M$ and wave function
$\Psi$. Yet if one closely examines the structure of the above equations, it can
be discovered that a mathematically similar problem was solved long ago by
Feshbach \cite{Feshbach} albeit in the rather different context of nuclear
reaction theory. Indeed there are a number of other contexts where analogous
problems have been solved, the case of mass and vertex renormalization in
pion-nucleon scattering being particularly noteworthy \cite{Afnan}. In the next
section we shall therefore use the method of Feshbach to derive the solution of
our covariant perturbation theory problem.

\subsection{Solution}
In this subsection we derive expressions for the bound state wave functions
$\Psi$, $\bPsi$, and the bound state mass $M$ corresponding to the full kernel
$K$ of \eq{K}.  Although our goal is to formulate the covariant perturbation
theory for this problem, we in fact derive expressions for $\Psi$, $\bPsi$,
and $M$, that are exact with all orders of $K_1$ being taken into account.
Starting from these exact expressions it is then trivial to generate all terms
of the perturbation series. To present our solution it will be convenient
to discuss the cases of nondegenerate and degenerate states,
separately.

\subsubsection{Nondegenerate case}

In the nondegenerate case, to each unperturbed bound state mass $M_u$ there
corresponds a unique bound state wave function $\Phi$. The unperturbed Green
function $G_u(P)$ then has the ``pole plus background'' structure, as given in
\eq{G_u_pole}. Having in mind that the full Green function $G(P)$ has a similar
structure as given in \eq{G_pole}, and that our goal is to relate the quantities
in these two expression, we begin by introducing a ``background'' Green function
$G^b(P)$ defined as the solution of the equation
\be 
G^b(P)=G^b_u(P)+G^b_u(P)K_1(P)G^b(P).           \eqn{G^b} 
\ee 
Note that $G^b(P)\ne \hat{G}^b(P)$ where $\hat{G}^b(P)$ was defined in
\eq{G_pole}. From \eq{G^b} it follows that 
\be 
(1+G^bK_1)^{-1}G^b = G_u^b, 
\ee 
where we have dropped the momentum arguments for convenience. Similarly \eq{GG_u} 
implies 
\be 
G(1+K_1G)^{-1} = G_u. 
\ee 
Subtracting the last two equations, we obtain 
\be 
G(1+K_1G)^{-1} - (1+G^bK_1)^{-1}G^b = \frac{i\Phi\bPhi}{P^2-M^2_u} 
\ee 
and therefore 
\be 
(1+G^bK_1)G - G^b(1+K_1G) = (1+G^bK_1)\frac{i\Phi\bPhi}{P^2-M^2_u}(1+K_1G). 
\ee 
Thus 
\be 
G  = G^b + \frac{(1+G^bK_1)i\Phi\bPhi(1+K_1G)}{P^2-M^2_u}, \eqn{GG} 
\ee 
which can be solved for $\bPhi(1+K_1G)$ by writing 
\be 
\bPhi(1+K_1G)  = \bPhi(1+K_1G^b) 
+ \frac{\bPhi K_1(1+G^bK_1)i\Phi\bPhi(1+K_1G)}{P^2-M^2_u}, 
\ee 
and then 
\be 
\bPhi(1+K_1G) 
= \left[1-\frac{i 
\bPhi(K_1+K_1G^bK_1)\Phi}{P^2-M^2_u}\right]^{-1}\bPhi(1+K_1G^b). 
\ee 
Using this result in \eq{GG} we obtain the result we are seeking: 
\be 
G(P)=\frac{i\psi(P)\bpsi(P)} 
{P^2-M^2_u-i\bPhi(P)\left[K_1(P)+K_1(P)G^b(P)K_1(P)\right]\Phi(P)}+G^b(P), 
\eqn{G_pole2} 
\ee 
where the functions $\psi(P)$ and $\bpsi(P)$ are defined by 
\be 
\psi(P) = \left[1+G^b(P)K_1(P)\right]\Phi(P) \eqn{psi}
\ee 
and 
\be 
\bpsi(P) = \bPhi(P)\left[1+K_1(P)G^b(P)\right]. \eqn{bpsi} 
\ee 
A comparison of \eq{G_pole2} with \eq{G_pole} shows that
$\Psi(\bP)=\sqrt{Z}\psi(\bP)$, and $\bPsi(\bP)=\sqrt{Z}\bpsi(\bP)$, where
\be 
Z = \left.\frac{1}{1-i\left\{\bPhi(P)\left[
K_1(P)+K_1(P){G^b}(P)K_1(P)\right]\Phi(P)\right\}'}\,\right|_{P^2=\bP^2=M^2},
\eqn{Z} 
\ee 
with the prime indicating a derivative with respect to $P^2$,
and 
\be 
M^2=M_u^2+ i\bPhi(\bP)\left[ K_1(\bP) 
+ K_1(\bP) G^b(\bP) K_1(\bP)\right]\Phi(\bP). \eqn{Mdiff} 
\ee 
In this respect it is worth noting that because all our wave functions and Green
functions are Lorentz covariant, the quantity in the curly brackets of \eq{Z}
[which also appears in \eq{Mdiff}], is a Lorentz scalar depending only on
$P^2$.

%It may be worth noting that \eq{Z} can be written in the alternative form 
%\be 
%Z = \frac{1}
%{1-i\bpsi(\bP)\left[K'_1(\bP)+K_1(\bP){G_u^b}'(\bP)K_1(\bP)\right]\psi(\bP)}. 
%\ee
Thus, in the nondegenerate case, 
the properly normalized wave functions for the full perturbation theory are 
\bea 
\Psi(\bP) &=& 
\left\{1-i\left\{\bPhi(\bP)\left[K_1(\bP)+
K_1(\bP){G^b}(\bP)K_1(\bP)\right]\Phi(\bP)\right\}'\right\}^{-1/2} 
[1+G^b(\bP)K_1(\bP)]\Phi(\bP),        \eqn{Psi} \\[2mm] 
\bPsi(\bP) &=& \bPhi(\bP)[1+K_1(\bP)G^b(\bP)] 
\left\{1-i\left\{\bPhi(\bP)\left[K_1(\bP)+
K_1(\bP){G^b}(\bP)K_1(\bP)\right]\Phi(\bP)\right\}'\right\}^{-1/2}. 
\eqn{bPsi} 
\eea 
We note that these wave functions satisfy the normalization condition 
\be
\left.
i\bPsi(P) \frac{\partial G^{-1}(P)}{\partial P^2}\Psi(P)
\right|_{P=\bP}=1.\eqn{Psi_norm} 
\ee 

\subsubsection{Reference frame dependence of the wave functions} 

As far as we know, all previous attempts at developing perturbation theory for
relativistic systems have considered bound states only at rest (see
e.g. \cite{Lepage}).  On the other hand, for observables involving scattering
off the bound state (e.g.\ electromagnetic form-factors) taking into account the
total momentum dependence of the bound state wave function is important. In the
relativistic case there are some subtleties in the determination of this
dependence perturbatively and at the same time in a manifestly covariant way.
One possible way to do this is to derive the wave function to the needed order
in the rest reference frame, and then to boost it in order to give it the
desired momentum. There are two disadvantages to this approach: one is that it
involves two separate steps - the perturbation expansion and the boosting. The
second disadvantage is that the unit vector $n=\bP/M$
which determines the boost \cite{Gasiorowicz}, itself may need to be calculated
perturbatively. To illustrate this, we consider the determination of a scalar
bound state
wave function $\Psi(\bP)$ to first  order in the perturbation. Showing
explicitly one relative momentum $p$ in addition to the total on-shell momentum
$\bP$, we first write the perturbed wave function as a boosted wave function at
rest:
\be 
\Psi(\bP,p)=S_{L_n}\Psi(L_n\bP,L_n p)=S_{L_n}\Psi_0(L_n p) 
\ee   
where $L_n$ is the boost Lorentz transformation, $L_n \bP=(M, {\bf 0})$, $S_{L_n}$ is
the associated transformation matrix acting on the spin indices of the constituents, and $\Psi_0(q)$ is the bound state
wave function at rest. Next step is to calculate $\Psi_0(q)$ to first order in
the perturbation: $\Psi_0(q)=(1+\eta_1)\Phi_0(q)$, where the first-order
correction factor $\eta_1$ is given explicitly in \eq{eta1}. Thus
\be 
\Psi(\bP,p)=S_{L_n}(1+\eta_1)\Phi_0(L_n p). \eqn{boost} 
\ee
As $L_n$ is a function of the unit vector $n=\bP/M=(\sqrt{{\bf P}^2+M^2}, {\bf
P})/M$, and therefore of $M$, and because we need $\Psi(\bP,p)$ up to first
order, the mass $M$, should be approximated up to first order in the
perturbation. Denoting the first order perturbation correction to $M^2$ by
$\delta_1$ [given explicitly in \eq{delta1}], the approximation $n(M)\approx
n(M_u+\delta_1/2M_u)$ should thus be used in \eq{boost} with a subsequent
expansion of the resulting $\Psi(\bP,p)$ up to first order in $\delta_1$. If
admixtures of higher-order corrections were acceptable, then this last expansion
could be neglected.

In what follows we show a more straightforward way to obtain the perturbed wave function $\Psi(\bP)$ when ${\bf P}\ne 0$.
For this purpose we shall require $G^b(P)$, which determines
the wave function via \eq{Psi}, to be Lorentz covariant; that is, we would like
it to transform kinematically under any Lorentz transformation $L$ of the
momenta involved, as $G^b(P;p',p)=S_L G^b(LP;Lp',Lp)S_L^\dagger$ where $p$ and
$p'$ are the initial and final relative momenta.
In order for this to be satisfied, $G_u^b(P;p',p)$ should also be Lorentz
covariant in view of \eq{G^b}. Using the definition (\ref{G_u_pole}) for
$G_u^b(P;p',p)$ one can see that the unperturbed wave function $\Phi(P,p)$
should also be a Lorentz covariant function under any Lorentz transformation $L$
of $P$ and $p$.  

Thus the essential problem is a practical one: how to construct a wave function
$\Phi(P,p)$ that is Lorentz covariant, and which satisfies the bound state
equation [first of \eqs{Phi}] for any $P$ such that $P^2=M_u^2$.  For this
purpose it is useful to have a separate notation for the bound state wave
functions, so to this end we denote by $\tPhi(\bP_u,p)$ all the solutions of the
bound state equation [first of \eqs{Phi}] for which the total momentum $\bP_u$
has the property $\bP_u^2=M_u^2$.  We then note that one cannot simply define
$\Phi(P,p)=\tPhi(\bP_u,p)$ where $P=(P_0,\bfP)$ and
$\bP_u=(\sqrt{\bfP^2+M_u^2},\bfP)$, so that $\Phi(P,p)$ does not depend on $P_0$
- such a $\Phi(P,p)$ cannot be Lorentz covariant since a Lorentz transformation
will change this function to $S_L\Phi(LP,Lp)$ which will necessarily depend on
the (arbitrary) value of $P_0$ (the three-vector part of $LP$ depends on $P_0$).

To make progress, we  note that the bound state wave function $\tPhi(\bP_u,p)$ is covariant under the transformation $\bP_u\rightarrow L\bP_u$, $p\rightarrow Lp$:
\be
\tPhi(\bP_u,p) = S_L\tPhi(L\bP_u,Lp).
\ee
As this is true for {\em any} four-vector $\bP_u$ satisfying $\bP_u^2=M_u^2$,
it will certainly be true for the four-vector  $M_u P/\sqrt{P^2}$
where $P$ is arbitrary.
Thus, if we define wave function $\Phi(P,p)$ as
\be
\Phi(P,p) = \tPhi\left(\frac{M_u P}{\sqrt{P^2}},p\right),\eqn{Phi_cov}
\ee
it immediately follows that
\be
\Phi(P,p) = S_L\Phi(LP,Lp), \eqn{good}
\ee
which is the statement that wave function $\Phi(P,p)$ is Lorentz covariant
in the way we need.  In
this way we have constructed a wave function $\Phi(P,p)$ that satisfies the
sought-after Lorentz covariance, while at the same time reducing to the bound
state wave function $\tPhi(\bP_u,p)$ as $P\rightarrow \bP_u$ (in fact
$\Phi(P,p)$, as defined by \eq{Phi_cov}, is the bound state wave function with
total momentum $M_u P/\sqrt{P^2}$). By choosing the form of $\Phi(P)$ given in
\eq{Phi_cov}, we guarantee that \eq{G_u_pole} is expressed in a manifestly
covariant way.  The immediate consequence of this is that the exact wave
function $\Psi(P)$ is given, up to a scalar normalization,
in a manifestly covariant way by \eq{psi}, and so is
each term in \eq{eta} corresponding to any order of perturbation theory for
$\Psi$. The same is valid for the denominator of \eq{G_pole2}, \eq{Mdiff} for
the mass, and the expression for the renormalization constant (\ref{Z}). If
instead we had chosen $\Phi(P)$ to transform differently from \eq{good}, even the fact that
the solution of \eq{Mdiff} does not depend on ${\bf P}$ would be hidden.

\subsubsection{Degenerate case}

In the degenerate case there is more than one solution $\Phi$ of the unperturbed
bound state equation, \eq{Phi}, for a single unperturbed bound state mass
$M_u$. Assuming an $r$-fold degeneracy, we denote such wave function solutions
as $\Phi_j$ where $j=1,2,3,\ldots, r$. In this case the pole structure of the
unperturbed Green function $G_u(P)$ is easily seen to be
\be
G_u(P) = \frac{i\sum_j\Phi_j(P)\bPhi_j(P)}{P^2-M_u^2}+G_u^b(P). \eqn{G_u_deg}
\ee
As for the non-degenerate case, we shall assume our wave functions to be
covariant but not dependent on $P^2$.
The wave functions $\Phi_j$ are, by the assumption of $r$-fold degeneracy,
linearly independent. Applying this fact to the pole structure of the identity
$G_uG_u^{-1}G_u=G_u$, we obtain the normalization condition for these wave
functions:
\be
\left.i\bPhi_i\frac{\partial G_u^{-1}(P)}{\partial P^2} \Phi_j\right|_{P=\bP_u}
 = \delta_{ij}.
\ee
\eq{G_u_deg} can be written exactly as \eq{G_u_pole} with $\Phi$ now defined
to be a row matrix whose elements are the $\Phi_j$:
\be
\Phi \equiv \left(\begin{array}{lllll}\Phi_1 & \Phi_2 & \Phi_3 & \ldots
& \Phi_r\end{array}\right),
\ee
with $\bPhi$ being the corresponding column matrix with elements $\bPhi_j$.
With this redefinition of $\Phi$ and $\bPhi$, the above derivation for the
nondegenerate case remains valid up until and including \eq{bpsi}. In this
way we obtain, for the degenerate case, that
\be
G(P)  = i\psi(P)A^{-1}(P)\bpsi(P) + G^b(P),
\eqn{G_pole2_deg}
\ee
where $\psi$ and $\bpsi$ are row and column matrices defined by elements
\be
\psi_j(P)=\left[1+G^b(P)K_1(P)\right]\Phi_j(P)\hspace{5mm}\mbox{and}\hspace{5mm}
\bpsi_j(P)=\bPhi_j(P)\left[1+K_1(P)G^b(P)\right],
\ee
respectively, and $A$ is an $r\times r$ matrix whose elements are
\be
A_{ij}(P) = (P^2-M_u^2)\delta_{ij}
- i\bPhi_i(P)\left[K_1(P)+K_1(P)G^b(P)K_1(P)\right]\Phi_j(P).
\ee

We are interested in the masses $M$ for which the Green function $G(P)$ of
\eq{G_pole2_deg} develops a bound state or resonance pole. This will happen
when the determinant of matrix $A(P)$ becomes zero. This, in turn, can be
determined by finding the matrix $S(P)$ which diagonalizes $A(P)$. With $S(P)$
determined, we have that
\be
D(P) \equiv S^{-1}(P) A(P) S(P) = \left(
\begin{array}{ccccc}
D_1(P) & 0 & 0 &\cdots & 0\\[2mm]
0 & D_2(P) & 0 & \cdots & 0 \\[2mm]
0 & 0 & D_3(P) & \cdots & 0 \\
\vdots & \vdots & \vdots & \ddots & \vdots \\
0 & 0 & 0 & \cdots & D_r(P)
\end{array}\right),
\ee
and
\be
G(P) = i\psi^S(P) D^{-1}(P) \bpsi^S(P) + G^b(P) \eqn{G_diag}
\ee
where
\be
D_{ij}(P) = (P^2-M_u^2)\delta_{ij}
-i\bPhi_i^S(P)\left[K_1(P)+K_1(P)G^b(P)K_1(P)\right]\Phi_j^S(P),
\ee
and
\be
\psi^S(P)\equiv \psi(P)S(P),\hspace{1cm}
\bpsi^S(P)\equiv S^{-1}(P)\bpsi(P),
\ee
with similar definitions holding for $\Phi^S(P)$ and $\bPhi^S(P)$.  Since
$\mbox{det} D(P) = \prod_jD_j(P) = 0$, the Green function $G(P)$ will have poles
at $P^2=M^2_j$, $j=1,2,3,\ldots, r$, where $M_j$ is the solution of the equation
\be
M_j^2=M_u^2
+i\bPhi^{S}_j(P_j)\left[K_1(P_j)+K_1(P_j)G^b(P_j)K_1(P_j)\right]\Phi_j^{S}(P_j),
\eqn{Mdiff_deg}
\ee
$P_j$ being any momentum satisfying $P_j^2=M_j^2$, and the functions
$\bPhi_j^{S}(P)$ and $\Phi_j^{S}(P)$ being the $j$'th elements of
$\bPhi^S(P)$ and $\Phi^S(P)$, respectively.

Taking into account the diagonal nature of $D(P)$, \eq{G_diag} can be written as
\be
G(P) = i\sum_j\psi_j^{S}(P) D_j^{-1}(P) \bpsi_j^{S}(P) + G^b(P).
\ee
Thus, assuming that the perturbed bound state mass $M_j$ is itself
nondegenerate, we can find its corresponding  wave function $\Psi_j$ as in the
nondegenerate case above: $\Psi_j=\sqrt{Z_j}\psi_j^{S}(P_j)$, where
\be
Z_j = \frac{1}
{1-i\left\{\bPhi_j^{S}(P_j)
\left[K_1(P_j)+K_1(P_j){G^b}(P_j)K_1(P_j)\right]\Phi_j^{S}(P_j)\right\}'}.
\ee
Thus, in the degenerate case of the unperturbed theory, the properly normalized
wave functions corresponding to the (nondegenerate) bound state mass $M_j$ of
the full perturbation theory, are
\bea
\Psi_j &=&\sqrt{Z_j}\, 
[1+G^b(P_j)K_1(P_j)]\,\Phi_j^{S}(P_j),  \eqn{Psi_deg}    \\[2mm]
\bPsi_j &=& \sqrt{Z_j}\, \bPhi_j^{S}(P_j)\,[1+K_1(P_j)G^b(P_j)].
\eea

\subsubsection{Comments}

The main results of this subsection are the expressions for $M^2$ and $\Psi$
given in the nondegenerate case by \eq{Mdiff} and \eq{Psi}, and in the
degenerate case by \eq{Mdiff_deg} and \eq{Psi_deg}, respectively. Not only are
these expressions exact and compact, but they can also be easily used to write
down the explicit perturbation series for these quantities. 
For this purpose it is most convenient to treat all functions of $P$ as functions
of $P^2$ and the unit four-vector $n=P/\sqrt{P^2}$, and at the same time to
use the covariant form for the unperturbed
wave function given by \eq{Phi_cov}, as then $\Phi$ will not depend on $P^2$.
For example, in the
nondegenerate case, to generate the perturbation series for $M^2$ we use
\eq{G^b} to write \eq{Mdiff} as an infinite series
\be
M^2 =  M_u^2 + i\bPhi\left[ \tK_1
+ \tK_1 \tG_u^b \tK_1+ \tK_1 \tG_u^b \tK_1 \tG_u^b \tK_1
+ \tK_1 \tG_u^b \tK_1 \tG_u^b \tK_1\tG_u^b \tK_1
+\ldots \right]\Phi
\ee
where a tilde over $K_1$ or $G_u^b$ indicates that this quantity is evaluated at
$P^2=M^2$. By making Taylor series expansions
\bea
\tK_1 &=& K_1+\delta K'_1+\frac{\delta^2}{2!}K''_1
+\ldots\\[3mm]
\tG_u^b &=& G_u^b+\delta G_u^{b\, '}
+\frac{\delta^2}{2!}G_u^{b\,''}+\ldots
\eea
where 
\be
\delta \equiv M^2-M_u^2
\ee
and each term without a tilde is evaluated at $P^2=M_u^2$, we can immediately
write $M^2$ as a perturbation series with respect to orders of $K_1\equiv
K_1(M_u)$:
\be
M^2 = M_u^2 + \delta_1 + \delta_2 + \delta_3 + \ldots
\ee
where
\bea
\delta_1 &=& i\bPhi K_1 \Phi \eqn{delta1} \\[2mm]
\delta_2 &=& i\bPhi\left[\delta_1 K'_1 +
K_1 G_u^b K_1\right] \Phi\\[2mm]
\delta_3 &=& i\bPhi\left[\delta_2 K'_1
+\frac{\delta_1^2}{2}K''_1
+ \delta_1\left( K_1 G_u^{b} K_1\right)'
+K_1 G_u^bK_1 G_u^bK_1\right] \Phi\\[2mm]
&\mbox{etc.}&\nonumber
\eea
Similarly, the wave function of \eq{Psi} can be written
as a perturbation series in orders of $K_1$:
\be
\Psi = \left(1+\eta_1+\eta_2+\eta_3 + \ldots   \eqn{eta}
\right)\Phi
\ee
where
\bea
\eta_1 &=& \textstyle\frac{1}{2}\Delta_1 + G_u^b K_1  \eqn{eta1}
\\[2mm]
\eta_2 &=& \textstyle\frac{1}{2}\Delta_2
+\frac{3}{8}\Delta_1^2
+\delta_1\left(G^b_u K_1\right)' + \eta_1 G_u^b K_1 \\[3mm]
\eta_3 &=& \textstyle\frac{1}{2}\Delta_3
+\frac{3}{4}\Delta_1\Delta_2+\frac{15}{48}\Delta_1^3
+\left(\delta_2 + \delta_1 \eta_1\right) \left(G_u^b K_1\right)'
+\frac{1}{2}\delta_1^2\left(G^b_u K_1\right)'' + \eta_2 G_u^b K_1\\[3mm]
&\mbox{etc.}&\nonumber
\eea
where $\Delta_i$ is derived from $\delta_i$ by putting an extra
derivative on each $K_1$ and $G_u^b$; that is,
\bea
\Delta_1 &=& i\bPhi K_1' \Phi\\[2mm]
\Delta_2 &=& i\bPhi\left[\delta_1 K''_1 +
(K_1 G_u^b K_1)'\right] \Phi\\[2mm]
\Delta_3 &=& i\bPhi\left[\delta_2 K''_1
+\frac{\delta_1^2}{2}K'''_1
+ \delta_1\left( K_1 G_u^{b} K_1\right)''
+(K_1 G_u^bK_1 G_u^bK_1)'\right] \Phi\\[2mm]
&\mbox{etc.}&\nonumber
\eea
A similar procedure can be used to generate the perturbation series for the
degenerate case.

It is worth noting that the perturbative corrections to the bound state wave
function, as derived here, are particularly important to take into account when
calculating corrections to vertices (electromagnetic, axial, etc.) within
constituent models. It is only by taking into account the appropriate order of
wave function perturbation exactly, will symmetry properties, like for example
gauge invariance, be preserved at each order in the vertex correction -- for a
concrete example, see Ref.\ \cite{njl} where \eq{eta1} was used to determine the
full lowest order pionic correction to the nucleon vertex function in the NJL
model.

It is also worth pointing out that in the case where the perturbation $K_1$ is
too large for a perturbative treatment, our expressions of \eq{Mdiff}, \eq{Psi}
\eq{Mdiff_deg}, and \eq{Psi_deg} may still be useful for performing practical
nonperturbative calculations of $M^2$ and $\Psi$. Indeed, in both the degenerate
and nondegenerate cases, the main calculational effort would be in solving
\eq{G^b} for the ``background'' Green function $G^b$. Yet this is an especially
simple equation, of standard Lippmann-Schwinger form, where $G^b$ has no pole at
$P^2=M^2$ and $G_u^b$ has no pole at $P^2=M_u^2$ (since they have been
subtracted), and where there are no singularities in the integration over
momenta. Even in the
unlikely event that $G_u^b$ happens to have an unsubtracted pole close to
$P^2=M^2$, this case can be easily handled numerically. Finally, it is useful
to note that $G_u^b$ has already been constructed for the important case of the
nonrelativistic Coulomb problem by Schwinger \cite{Schwinger} -- a result that
can be easily adapted to the relativistic Coulomb case \cite{Ivanov}.  

\section{Discussion and Summary}

In this work we have presented a general formulation of perturbation theory
applicable to bound states and resonances where the bound state equations
involve kernels and inverse free Green functions that have an arbitrary energy
dependence. Our formulation is thus directly applicable to the important case of
relativistic quantum field theory. One can consider our results as extending the
well-known time-independent perturbation theory of quantum mechanics to the case
where the kernels are energy-dependent and where the inverse propagators are
non-linear in the energy.

In particular, we have derived expressions for the bound state (or resonance)
mass $M$ and wave function $\Psi$ of a system whose interaction kernel $K$
consists of a part $K_0$ for which the corresponding Green function $G_u$ is
known, and a part $K_1$ which plays the role of a perturbation. Our results for
$M$ and $\Psi$ are contained in \eq{Mdiff} and \eq{Psi} for the nondegenerate
case, and in \eq{Mdiff_deg} and \eq{Psi_deg} for the degenerate case, and have
the feature that they are exact, with the perturbation $K_1$ taken into account
to all orders. The key element in these expressions is the Green function $G^b$
which needs to be found by solving \eq{G^b}.  For sufficiently small $K_1$,
\eq{G^b} can be solved simply by iteration, in this way generating a
perturbation expansion in $K_1$ that is the analogue of the time-independent
perturbation theory of quantum mechanics. On the other hand, if $K_1$ is not
small enough to generate a convergent perturbation series, \eq{G^b} could still
be solved by standard numerical techniques for integral equations.

As far as we know, our formulation of the perturbation theory problem
is new. However, there are a few alternative formulations available in the
literature, all presented for the particular case of relativistic quantum field
theory. The first of these is a method where the perturbation series for
$M^2$ and $\Psi$ are expressed in terms of contour integrals. Originally
developed by Kato \cite{Kato} and described in Messiah's standard text
\cite{Messiah} for the case of quantum mechanics, the contour method was
extended to the covariant case by Lepage \cite{Lepage} and used, for example, by
Murato \cite{Murota}.  Another method, due to Bodwin and Yennie \cite{Bodwin},
is closest in spirit to our approach, but does not have the feature of having
closed expressions for the perturbed mass and wave function.  A third approach
is the recent formulation of Ivanov {\em et al.}\ \cite{Ivanov} whose
perturbative expansion is expressed in terms of a certain ``relativistic
generalization of a projection operator''.  In this approach the second
derivative of the inverse free propagator, $\partial^2 G_0^{-1}/\partial E^2$,
looks very much like a genuine and necessary relativistic feature, yet it does
not appear in our formulation at all and is thus just an artifact of the
particular derivation used. Similarly, the expression for the lowest-order wave
function correction derived directly from Eq.\ (9) of Ref.\ \cite{Ivanov}
contains four terms against our only one. 

In each of the above three alternative approaches, perturbative corrections
to the bound state wave function were derived only for the special case where the
bound state is at rest. Thus, in order to describe scattering process where the bound
state has non-zero total momentum, such wave function corrections need to be
modified by the appropriate Lorentz boost (that itself depends on the order of
perturbation being considered). By contrast, our approach has enabled us to
write expressions for the bound state wave function corrections that are Lorentz
covariant at each order of the perturbation, thus avoiding the step of boosting from
the rest frame. Although all perturbation expansions
must mathematically be identical, it is evident that the expressions provided by
our \eq{Mdiff}, \eq{Psi}, \eq{Mdiff_deg}, and \eq{Psi_deg} are the simplest both
practically and conceptually.

\acknowledgments

We would like to thank A.\ G.\ Rustesky for fruitful discussions.
This work is partially supported by the Engineering and Physical Sciences
Research Council (U.K.).

%\appendix
%\section*{}

% now the references. delete or change fake bibitem. delete next three
%   lines and directly read in your .bbl file if you use bibtex.

% \begin{figure}
% \caption{}
% \label{}
% \end{figure}

% \begin{table}
% \caption{}
% \label{}
% \end{tabular}
% \end{table}
\end{document}